\documentstyle[aps,preprint,tighten]{revtex}
\def\tr{\text{tr}}
\def\x{{\bf x}}
\def\y{{\bf y}}

\def\L{{\text{L}}}
\def\R{{\text{R}}}
\def\LR{{\text{L,R}}}

\def\Nf{N_{\text{f}}}

\def\d{\partial}
\def\<{\langle}
\def\>{\rangle}

\begin{document}
%\twocolumn[\hsize\textwidth\columnwidth\hsize\csname@twocolumnfalse\endcsname

\preprint{CU-TP-960}
\title{Hydrodynamics of Nuclear Matter in the Chiral Limit}

\author{D.T.~Son}

\address{Physics Department, Columbia University, New York, New York 10027}
\address{and RIKEN-BNL Research Center, Brookhaven National Laboratory,
Upton, New York 11973}

\date{December 1999}

\maketitle

\begin{abstract}
Using the Poisson bracket method, we construct the hydrodynamics of
nuclear matter in the chiral limit, which describes the dynamics of
all low-energy degrees of freedom, including the fluid-dynamical and
pionic ones.  The hydrodynamic equations contain, beside five Euler
equations of relativistic fluid dynamics, $\Nf^2-1$ second order
equations describing propagating pions and $\Nf^2-1$ first order
equations describing the advection of the vector isospin
charges.  We present hydrodynamic arguments showing that the pion
velocity vanishes at the second order phase transition at $\Nf=2$.
\end{abstract}
%\vskip 2pc]

Hydrodynamics \cite{Forster} is the theory describing the
low-frequency, long-wavelength dynamics of liquids (or, in an extended
sense, of any system).  In this regime most degrees of freedom become
irrelevant since they relax during the time characteristic of particle
collisions; the only ones that survive are either those related to the
conservation laws or the phases of the order parameters of broken
continuous symmetries.  The simplest example is normal fluids, where
hydrodynamic variables arise due to the conservation of energy,
momentum, and particle number.  In superfluid He$^4$ and He$^3$,
additional hydrodynamic variables emerge from the symmetry breaking by
the condensate.  Although in all of these systems the physics is quite
complicated at the molecular level, at large scales the hydrodynamic
equations have simple forms dictated by the symmetries, the pattern of
symmetry breaking, and the conservation laws.  Such equations
typically involve unknown coefficients, which can be computed from the
microscopic theory or measured in experiments.

A similar philosophy is shared by the chiral perturbation theory,
which describes the long-distance dynamics of QCD with light
quarks\cite{Georgi}.  At low energies, QCD is a strongly coupled
theory where not much can be computed, at least at this moment, in a
reliable fashion.  However, well below the chiral scale (about 1 GeV),
the dynamics is determined by the chiral Lagrangian, which can be
written down knowing only the chiral symmetry and the pattern of
chiral symmetry breaking of QCD.  To the lowest order, pions are
governed by the nonlinear sigma model, the only free parameter of
which is the pion coupling constant $f_\pi$, which can be determined
by matching the predictions of theory with experiment.

In nuclear matter in the chiral limit, the low-energy degrees of
freedom include both the fluid-dynamical variables (the
energy-momentum tensor) and the chiral ones that describe the massless
Goldstone modes arising from the breaking of chiral symmetry.  All
these degrees of freedom are coupled to each other; therefore, a full
hydrodynamic treatment must include all these modes.  In this respect,
the hydrodynamics of nuclear matter is more similar to that of
superfluids than of normal fluids.  Treatments so far have largely
dealt with the fluid dynamical and chiral variables separately,
ignoring the interplay between the two \cite{Scavenius}.

The purpose of this paper is to construct the hydrodynamic theory of
nuclear matter in the chiral limit, capable of describing {\em all}
low-energy degrees of freedom of the latter.  The primary place where
such a theory can be applied is in the theory of heavy-ion collisions,
where pion modes can arise by thermal activation or from the spinodal
instabilities at the chiral phase transition.

{\em Degrees of freedom}.---For definiteness, we will be working in
$\Nf=2$ QCD, but most formulas remain valid at any value of $\Nf$.  As
the first approximation, we take the quark masses to be zero,
$m_u=m_d=0$.  The nonzero pion mass, hopefully, can be taken care of
by expanding around the chiral limit.  We will assume that the
temperature and the chemical potential are low enough so that the
chiral symmetry is broken.

The first step toward a hydrodynamic description is to identify the
hydrodynamic variables.  As mentioned above, the latter are the
densities of conserved charges or the phases of order parameters of
broken continuous symmetries.  In nuclear matter, different vacua
arising from chiral symmetry breaking are characterized by the phase
$\Sigma$ of the condensate, where $\Sigma\in$ SU(2), which transforms
under chiral rotations as $\Sigma\to L\Sigma R^\dagger$.  Slow
variations of $\Sigma$ correspond to soft pions, which, due to the
derivative coupling, relax much slower than a typical mode, and,
hence, are hydrodynamic degrees of freedom.  The conservation laws
include those of the energy and the momentum, as well as the baryon
number $n=\int\!d\x\,q^\dagger q$ ($q$ denotes quarks), and the left-
and right-handed isospin charges $\int\!d\x\,\rho^a_\LR(\x)=
\int\!d\x\,q_\LR^\dagger{\lambda^a\over2}q_\LR$, which are the
generators of left and right isospin rotations.  Therefore, the
hydrodynamics variables are the energy and momentum densities
$T^{0\nu}$, the baryon number density $n(\x)$, the left and right
isospin charge densities $\rho^a_\LR(\x)$, and the SU(2) phase of the
condensate $\Sigma(\x)$.

{\em Hydrodynamics of a reduced set of variables}.---A nontrivial
hydrodynamics can be already constructed for a reduced set of
variables, chosen here to contain $\rho^a_\LR$ and $\Sigma$.  In this
simplified treatment, we ignore the variation of the fluid dynamical
degrees of freedom, regarding the fluid as frozen.  Such a theory is
not the full theory, but its construction is much simpler and the
equations obtained are suggestive of those in full hydrodynamics,
which will be constructed later.  

The method we use to write down the hydrodynamic equations is the
elegant Poisson bracket technique \cite{DzyalVol,ZakhKuz}, which
regards the ideal hydrodynamics of any fluid as a Hamiltonian system
defined by a Hamiltonian and a set of Poisson brackets between
hydrodynamic variables.  The Poisson brackets arise from the canonical
commutators in the microscopic theory; the ones between $\rho_\LR$ are
the classical version of the current algebra commutators,
\begin{eqnarray}
  \{ \rho_\L^a(\x),\, \rho_\L^b(\y) \} & = & 
    -f^{abc}\rho_\L^c(\x)\delta(\x-\y), \nonumber\\
  \{ \rho_\R^a(\x),\, \rho_\R^b(\y) \} & = & 
    -f^{abc}\rho_\R^c(\x)\delta(\x-\y), \label{PBrhorho}\\
  \{ \rho_\L^a(\x),\, \rho_\R^b(\y) \}, & = 0 \nonumber
\end{eqnarray}
while those between $\rho_\LR$ with $\Sigma$ are defined by the
transformation laws of $\Sigma$ under chiral rotations, since
$\rho_\LR$ are the densities of charges that generate these
transformations,
\begin{eqnarray}
  \{ \rho_\L^a(\x),\, \Sigma(\y) \} & = & 
    -i{\lambda^a\over2} \Sigma(\x) \delta(\x-\y), \nonumber\\
  \{ \rho_\R^a(\x),\, \Sigma(\y) \} & = & 
     i\Sigma(\x){\lambda^a\over2}\delta(\x-\y). \label{PBrhoSigma}
\end{eqnarray}
Finally, the Poisson brackets of $\Sigma$ with itself vanishes, 
\begin{equation}
  \{ \Sigma(\x),\, \Sigma(\y) \} = 0.
  \label{PBSigmaSigma}
\end{equation}

Now let us turn to the Hamiltonian.  We will limit ourselves to the
leading order in derivatives of $\Sigma$, since we are interested in
the dynamics at the largest scales.  We will also keep only
leading-order terms in powers of $\rho$.  The most general form of the
Hamiltonian consistent with chiral symmetry is
\begin{equation}
  H  = \int\!d\x\, \biggl[ 
      {f_s^2\over4}\tr\d_i\Sigma^\dagger\d_i\Sigma + 
      {1\over f_t^2}\tr(\rho_\L-\Sigma\rho_\R\Sigma^\dagger)^2
     +{1\over f_v^2}\tr(\rho_\L+\Sigma\rho_\R\Sigma^\dagger)^2\biggr], 
   \label{Hamiltonian}
\end{equation}
where $\rho_\LR=\rho^a_\LR{\lambda^a\over2}$ and $f_s$, $f_t$, and
$f_v$ are constants with the dimension of mass, whose physical meaning
will become clear later.

Taking the Poisson brackets with the Hamiltonian (\ref{Hamiltonian}),
we obtain the equations of motion of the hydrodynamic variables,
\begin{eqnarray}
  \d_t\Sigma & = & - {2i\over f_t^2}(\rho_\L\Sigma - \Sigma \rho_\R),
    \label{eq1} \\
  \d_t\rho_\L & = & - \d_i J^\L_i,
    \label{eq3} \\
  \d_t\rho_\R & = & - \d_i J^\R_i,
    \label{eq4}
\end{eqnarray}
where 
\[
  J^\L_i = {i\over4}f_s^2\Sigma\d_i\Sigma^\dagger, \qquad 
  J^\R_i = {i\over4}f_s^2\Sigma^\dagger\d_i\Sigma . 
\]
Equations (\ref{eq3}) and (\ref{eq4}) reflects the conservation of
left- and right-handed flavor charges, where $J^\LR_i$ plays the role
of the chiral isospin currents.  Note that $f_v$ does not appear in
Eqs.\ (\ref{eq1})--(\ref{eq4}); the reason is that
$\tr(\rho_\L+\Sigma\rho_\R\Sigma^\dagger)^2=
\tr(\rho_\L^2+\rho_\R^2+2\rho_\L\Sigma\rho_\R\Sigma^\dagger)$ is a
Casimir operator.  Equations (\ref{eq1})--(\ref{eq4}) completely
determine the dynamics of $\rho$ and $\Sigma$.  These differential
equations are of first order in time and describe the evolution of
$3(\Nf^2-1)=9$ variables.

{\em The relation to the nonlinear sigma model}.---The hydrodynamic
equations (\ref{eq1})--(\ref{eq4}) should be contrasted with the field
equations of the nonlinear sigma model, which describes the dynamics
of pions at zero temperature and chemical potential.  The latter is
composed of three equations of second order in time derivative, which
can be rewritten as six first order equations describing the evolution
of $\Sigma$ and $\d_t\Sigma$.  Therefore our hydrodynamics contains at
least three extra degrees of freedom that are not presented in the
nonlinear sigma model.

To find the relation of the hydrodynamic equations with the field
equations of the nonlinear sigma model, one solves Eq.\ (\ref{eq1})
with respect to $\rho_\L$ and $\rho_\R$ and expresses them via
$\d_t\Sigma$ and a new variable $\alpha$,
\begin{eqnarray}
  \rho_\L & = & -{i\over4}f_t^2\Sigma\d_t\Sigma^\dagger + 
  {1\over2}\alpha,
  \label{qLdef}\\
  \rho_\R & = & -{i\over4}f_t^2\Sigma^\dagger\d_t\Sigma + 
     {1\over2}\Sigma^\dagger\alpha\Sigma.
  \label{qRdef}
\end{eqnarray}
In particular, $\alpha=\rho_\L+\Sigma\rho_\R\Sigma^\dagger$, therefore
in the vacuum where $\Sigma=1$, $\alpha$ is the density of isovector
charge.  The equation of motion for $\alpha$ reads
\begin{equation}
  \dot\alpha = -{1\over2}[\Sigma\d_t\Sigma^\dagger,\alpha],
  \label{alphadot}
\end{equation}
while the equation for $\Sigma$ is now second order in time derivative,
\begin{equation}
  if_t^2\d_t(\Sigma\d_t\Sigma^\dagger)
  -if_s^2\d_i(\Sigma\d_i\Sigma^\dagger) 
  + [\Sigma\d_t\Sigma^\dagger,\alpha] = 0.
  \label{Udot}
\end{equation}
Equation (\ref{alphadot}) always allows $\alpha=0$ as a solution.  In
this case, Eq.\ (\ref{Udot}) reduces to the field equation of the
nonlinear sigma model with pion velocity equal $v_\pi=f_s/f_t$.
Therefore $f_s$ and $f_t$ play the role of spatial and temporal pion
decay constants, respectively \cite{Pisarski}.  The nonlinear sigma
model can be interpreted as the Hamiltonian system
(\ref{PBrhorho})--(\ref{Hamiltonian}) with the {\em constraint}
$\rho_\L+\Sigma\rho_\R\Sigma^\dagger=0$.

In general, however, $\alpha$ needs not vanish. Equation
(\ref{alphadot}) implies that, in the dissipationless limit we are
considering, $\alpha$ only {\em precesses} with time.  Since Eq.\
(\ref{alphadot}) is first order in time derivative, once dissipation
is included into the theory $\alpha$ will become a true diffusive
mode.  In any case, when $\alpha\neq0$, our hydrodynamic equations are
different from the field equations of the nonlinear sigma model.

{\em Hydrodynamic pion condensation}.---To understand the physical
meaning of the term proportional to $\alpha$ in Eq.\ (\ref{Udot}),
consider the case where $\Sigma$ makes only small variations around
$\Sigma=1$.  As noted above, $\alpha$ is now the isovector charge.
Subsequently, $\alpha$ can be nonzero if {\em baryons} are included
into the theory.  For example, if the baryon background contains more
neutrons than protons, then $\alpha^3=\rho_p-\rho_n$ is nonzero and
negative.  It is clear from the discussion above that $\alpha$ is the
only baryonic degree of freedom that enter the hydrodynamic theory.

By expanding around $\Sigma=1$, using
$\Sigma=\exp(if_t^{-1}\lambda^a\pi^a)$, one finds the following
linearized equation for the pion field on an isospin-asymmetric
background:
\begin{equation}
  \d_0^2\pi^a - v_\pi^2\d_i\pi^a + f_t^{-2}f^{abc}\d_0\pi^b\alpha^c
  = 0,
  \label{linearpi}
\end{equation}
Such an equation can also be obtained from the mean-field
approximation of the chiral perturbation theory, by replacing in the
interaction Lagrangian $-{1\over2}f_\pi^{-2}f^{abc}\pi^a\d_\mu\pi^b
\bar{N}\gamma^\mu{\lambda^c\over2}N$ the isospin baryon density
$\bar{N}\gamma^0{\lambda^c\over2}N$ by its mean value $\alpha^c$.  We
have shown that this mean-field procedure becomes reliable in the
hydrodynamic limit, provided the in-medium pion decay constants $f_t$
and $f_s$ are used.  Equation (\ref{linearpi}) predicts a split between
the dispersion relations of $\pi^+$ and $\pi^-$ in neutron-rich
backgrounds where $\alpha^3<0$.  This is the hydrodynamic equivalence
of pion condensation \cite{picond}.

{\em Pion velocity near the second order phase transition}.---Let us
consider $\Nf=2$ and assume that the baryon chemical potential $\mu$
is zero or small enough so that the phase transition in temperature is
second order \cite{PisarskiWilczek}.  We will argue here that the pion
velocity vanishes at the critical temperature.  Indeed, as one
approaches the critical temperature $T_c$, the dependence of the
Hamiltonian (\ref{Hamiltonian}) on the phase of the condensate
$\Sigma$ should become weaker, and, at $T=T_c$, $H$ should not depend
explicitly on $\Sigma$.  The latter can happen only if at the critical
temperature $f_s=0$ and $f_t=f_v$.  Now since $f_v$ enters the
Hamiltonian (\ref{Hamiltonian}) as $\alpha^2/f_v^2$, it is related to
the response of nuclear matter in the {\em isovector} channel.  Such a
quantity has no reason to vanish at the phase transition.  Therefore
one can expect $f_v\neq0$ at $T_c$ and, hence, $f_t$ also does not
vanish at the critical temperature, in contrast to $f_s$.  Therefore
the pion velocity $v_\pi=f_s/f_t$ approaches zero when $T\to T_c$.

{\em Full hydrodynamics}.---Now let us turn to the discussion of the
full hydrodynamic equations, which contains not only chiral variables
but also the fluid dynamical ones.  These equations can also be
derived from the general Poisson bracket technique similar to the one
used above.  The hydrodynamic variables include, beside $\Sigma$ and
$\rho_\LR$, the baryon density $n(\x)$, the entropy density $s(\x)$,
and the momentum density $T^{0k}(\x)$.  The nonvanishing Poisson
brackets, beside those written in Eqs.\
(\ref{PBrhorho})--(\ref{PBSigmaSigma}), are
\begin{eqnarray*}
  \{T^{0i}(\x),\, A(\y)\} & = & A(\x)\d_i\delta(\x-\y),
     \qquad A=s,\, n,\, \rho_\LR, \\
  \{T^{0i}(\x),\, \Sigma(\y)\} & = & - \d_i\Sigma(\x)\delta(\x-\y), \\
  \{T^{0i}(\x),\, T^{0k}(\y)\} & = & 
     \biggl[ T^{0k}(\x){\d\over\d x^i} - T^{0i}(\y){\d\over\d y^k}
     \biggr] \delta(\x-\y).
\end{eqnarray*}
In particular, now $\tr(\rho_\L+\Sigma\rho_\R\Sigma)^2$ is no longer a
Casimir of the Poisson algebra: it has a nonzero Poisson bracket with
$T^{0k}$.  The Hamiltonian is chosen in the most general form
consistent with symmetries and containing only second order of
$\rho_\LR$ and derivatives of $\Sigma$,
\begin{eqnarray}
  H &=& \int\!d\x\, T^{00}(\x) = \int\!d\x\,
    \biggl(\epsilon + f_{ij}\tr\d_i\Sigma\d_j\Sigma^\dagger
      +{a\over 2}\tr(\rho_\L-\Sigma\rho_R\Sigma^\dagger)^2 +
    {a_v\over 2}\tr(\rho_\L+\Sigma\rho_R\Sigma^\dagger)^2\nonumber\\
    & & - ic_k\tr(\rho_\L\Sigma\d_k\Sigma^\dagger+
    \Sigma^\dagger\d_k\Sigma\rho_R)\biggr),
  \label{Hfull}
\end{eqnarray}
where $\epsilon$, $f_{ij}$, $a$, $a_v$, and $c_k$ are functions of
$s$, $n$ and $T^{0k}$.  It is, however, more convenient to work with
the conjugate variables: the temperature $T=\d T^{00}/\d s$, the
chemical potential $\mu=\d T^{00}/\d n$, and the velocity $v^k=\d
T^{00}/\d T^{0k}$.  In the frame where $v^k=0$, $\epsilon$ is
determined by the nuclear equation of state, while
$f_{ij}=\delta_{ij}f_s^2/4$, $a=2f_s^{-2}$, $a_v=2f_v^{-2}$ are
functions of the local temperature $T$ and chemical potential $\mu$,
and $c_k=0$.  The equations of motion for our variables can be found
by taking the Poisson brackets with the Hamiltonian (\ref{Hfull}).
The condition that $\d_0T^{00}=-\d_iT^{0i}$ allows one to determine
the velocity dependence of $\epsilon$, $f_{ij}$, $a$, $a_v$, and
$c_k$, which turns out to be equivalent to the condition of boost
invariance.  The latter allows the final equations to be written in a
relativistically covariant form, although our formalism is Hamiltonian
in nature.  After quite laborious, but straightforward, calculations
\cite{Son}, one finds that the full set of hydrodynamic equations
consists of (i) a second order equation for $\Sigma$,
\begin{equation}
  i\d_\mu[(f_t^2-f_s^2)u^\mu u^\nu\Sigma\d_\nu\Sigma^\dagger +
  f_s^2\Sigma\d^\mu\Sigma^\dagger]
  +[u^\mu\Sigma\d_\mu\Sigma^\dagger,\alpha] = 0;
  \label{dSigmafull}
\end{equation}
(ii) a first order equation describing the advection and precession of
the isovector charge $\alpha$,
\begin{equation}
  \d_\mu(u^\mu\alpha) = 
  - {1\over2}[u^\mu\Sigma\d_\mu\Sigma^\dagger, \alpha];
  \label{dalphafull}
\end{equation}
(iii) the continuity equation for the baryon charge,
\begin{equation}
  \d_\mu(u^\mu n) = 0;
  \label{dBfull}
\end{equation}
and (iv) the conservation of energy momentum,
\begin{equation}
  \d_\mu T^{\mu\nu} = 0,
  \label{dTfull}
\end{equation}
where the energy-momentum tensor $T^{\mu\nu}$ is a sum of a fluid
dynamical part and a field (pion) part,
\[
  T^{\mu\nu} = (\rho+p)u^\mu u^\nu - pg^{\mu\nu} +
  {f_s^2\over4}\tr(\d^\mu\Sigma\d^\nu\Sigma^\dagger +
  \d^\nu\Sigma\d^\mu\Sigma^\dagger).
\]
Moreover, both the energy density $\rho$ and the pressure $p$ receive
a contribution from the pion field $\Sigma$ and the density of
isovector charge $\alpha$.  In particular,
\begin{equation}
  p = p_0 + 
  {1\over4}(f_t^2-f_s^2)u^\mu u^\nu\tr\d_\mu\Sigma\d_\nu\Sigma^\dagger +
  {f_s^2\over4}\tr\d_\mu\Sigma\d^\mu\Sigma^\dagger
  +f_v^{-2}\tr\alpha^2,
  \label{pfull}
\end{equation}
where $p_0=p_0(T,\mu)$ is the pressure at $\Sigma=1$ and $\alpha=0$,
and $\rho$ is related to $p$ by Legendre transformation,
\begin{equation}
  \rho = \rho_0 + {1\over4}(\hat K +1)(f_t^2-f_s^2)u^\mu u^\nu 
  \tr\d_\mu\Sigma\d_\nu\Sigma^\dagger
  +(\hat K-1){f_s^2\over4}\tr\d_\mu\Sigma\d^\mu\Sigma^\dagger
  +f_v^{-4}(\hat K+1)f_v^2\tr\alpha^2,
  \label{rhofull}
\end{equation}
where $\hat K = T\d/\d T+\mu\d/\d\mu$.  As in relativistic fluid
dynamics, it can be shown that the conservation of entropy is a
consequence of energy-momentum conservation (\ref{dTfull}), baryon
number conservation (\ref{dBfull}), and thermodynamic relations.  The
conservation of left- and right-handed isospin currents,
\begin{eqnarray*}
  J_\L^\mu &=& -{i\over4}
    [(f_t^2-f_s^2)u^\mu u^\nu\Sigma\d_\nu\Sigma^\dagger+
    f_s^2\Sigma\d^\mu\Sigma^\dagger]+{1\over2}u^\mu\alpha, \\
  J_\R^\mu &=& -{i\over4}
    [(f_t^2-f_s^2)u^\mu u^\nu\Sigma^\dagger\d_\nu\Sigma+
    f_s^2\Sigma^\dagger\d^\mu\Sigma]+
    {1\over2}u^\mu\Sigma^\dagger\alpha\Sigma,
\end{eqnarray*}
follows from Eqs.\ (\ref{dSigmafull}) and (\ref{dalphafull}).

Equations (\ref{dSigmafull}) and (\ref{dalphafull}) are the
straightforward generalization of Eqs.\ (\ref{alphadot}) and
(\ref{Udot}).  At small temperatures and chemical potentials,
$f_t\approx f_s$ and $\alpha\approx0$, the equation of motion for
$\Sigma$ decouples from the fluid flow and coincides with the field
equation of the nonlinear sigma model.  When $\Sigma=1$, the
hydrodynamic equations reduce to those of relativistic fluid dynamics.
Thus, our equations include both the relativistic fluid dynamics and
the nonlinear sigma model as special cases.  This makes our equations
ideal for the study of the evolution of disoriented chiral condensates
\cite{DCC} and any other situation where soft pions are in multiply
occupied states.  For example, the question of coupling between sound
waves and pions can be addressed using these equations.

The treatment of this paper could be improved in several directions.
First, one can include the effects of quark masses by introducing the
mass term $\tr(M\Sigma)+\text{h.c.}$ into the Hamiltonian, where $M$
depends on $T$ and $\mu$.  Because of the relatively large pion mass,
such a term could be quite important.  This will modify Eq.\
(\ref{dSigmafull}) for $\Sigma$ and give additional contributions to
the pressure and the energy density in Eqs.\ (\ref{pfull}) and
(\ref{rhofull}), but the main features of the theory remain unchanged
\cite{Son}.  Other next-to-leading terms in the chiral expansion can
be also added.  Second, the dissipative effects can be introduced,
although one will certainly encounter the well-known problems of
ambiguity and instability of viscous relativistic fluid dynamics
\cite{HiscockLindblom}.

In conclusion, let us notice that the method developed in this paper
should be applicable, with some modifications, to the
color-flavor-locking phase of finite-density QCD with $\Nf=3$
\cite{CFL}.  The only difference is the breaking of U(1) baryon
symmetry.  The hydrodynamics contains one additional hydrodynamic
variable [the $\text{U(1)}_{\text{B}}$ phase], which gives rise to the
superfluid baryon number current.

The author thanks Misha Chertkov, David Kaplan, Roman Jackiw, Rob
Pisarski, Krish\-na Rajagopal, Dirk Rischke, Misha Stephanov, and Larry
Yaffe for fruitful discussions.  He thanks RIKEN, Brookhaven National
Laboratory and the U.S. Department of Energy [DE-AC02-98CH10886] for
providing the facilities essential for the completion of this work.
This work was supported, in part, by DOE Grant No.\ DE/FG02-92ER40699.

\end{document}